\documentclass{article}%
\usepackage{amsmath}
\usepackage{amsfonts}
\usepackage{amssymb}
\usepackage{graphicx}
\usepackage{float}
\usepackage{subcaption}
\usepackage{multirow}
\usepackage{color}
\usepackage{longtable}
\usepackage[dvipsnames]{xcolor}
\usepackage[bookmarks=true,colorlinks=true]{hyperref}
\hypersetup{linkcolor=blue}
\hypersetup{citecolor=blue}
\hypersetup{    
	urlcolor=blue,
}
\newcommand{\bpartial}{\mathop{\partial\kern -4pt\raisebox{.8pt}{$|$}}}
\newcommand{\bra}{\mathopen{[\kern-1.6pt[}}
\newcommand{\ket}{\mathclose{]\kern-1.5pt]}}
\newcommand{\bbra}{\mathopen{[\kern-2.2pt[\kern-2.3pt[}}
\newcommand{\bket}{\mathclose{]\kern-2.1pt]\kern-2.3pt]}}

\makeindex
\date{}

\begin{document}
	\title {\large{ \bf 
		Quasinormal modes of three $(2+1)$-dimensional 
			 black holes in string theory, conformal gravity, and  Hu-Sawicki $F(R)$ theory  via the Heun function
			 } }
	
	\vspace{3mm}
		\author{  \small{ \bf  F. Naderi }\hspace{-1mm}{ \footnote{
					 e-mail:
			 f.naderi@azaruniv.ac.ir (Corresponding author)}} ,{ \small	} \small{ \bf  A. Rezaei-Aghdam}\hspace{-1mm}{
		\footnote{
			 e-mail:	rezaei-a@azaruniv.ac.ir }} \\
		 		{\small{\em
			Department of Physics, Faculty of Basic Sciences, Azarbaijan Shahid Madani University}}\\
	{\small{\em   53714-161, Tabriz, Iran  }}}

\maketitle

\begin{abstract}
    
 	We study the propagation of massless fermionic fields, implementing a family of special functions: Heun functions, in solving the wave equation in three three-dimensional backgrounds, including the BTZ black hole in string theory and   Lifshitz black hole solutions in conformal gravity and  Hu-Sawicki $F(R)$ theory. The main properties of the selected black hole solutions is that their line elements are Weyl related to that of a homogeneous spacetime, whose spatial part possesses Lie symmetry, described by Lobachevsky-type geometry with arbitrary negative Gaussian curvature.  Using the Weyl symmetry of massless Dirac action, we consider the perturbation equations of fermionic fields in relation to those of the  homogeneous background, which having definite singularities,  are transformed into Heun equation. We point out the existence of quasinormal modes labeled by the accessory parameter of the Heun function.
 The distribution of the quasinormal modes has been clarified to satisfy the boundary conditions that require ingoing and decaying waves at the event horizon and conformal infinity, respectively. It turned out that the procedure based on the Heun function, beside reproducing the previously known results obtained via hypergemetric function for the BTZ and Lifshitz black hole solution in conformal gravity,  brings up new families of quasinormal frequencies, which can also contain purely imaginary modes.  Also, the analysis of the quasinormal modes shows that with the negative imaginary part of complex frequencies $\omega=\omega_{Re}+i\omega_{Im}$, the fermionic perturbations are stable in this background.

\end{abstract}
\section{Introduction}
The $(2+1)$-dimensional models of gravity,  initially considered in   \cite{staruszkiewicz1963gravitation}, have attracted attention in recent years.
Apart from being considered a toy model of quantum gravity to
survey the classical and quantum dynamics of point sources \cite{deser1984three,deser1988classical,t1988non}, 
these models have been used in the representation of Chern-Simons theory for $(2+1)$-dimensional gravity   \cite{WITTEN1989113,WITTEN19,witten1991quantization}.  
Classical and quantum solutions to $(2+1)$-dimensional gravity theories have been widely investigated, for instance, in \cite{21Hamber2012,21Carlip1994,PhysRevD.90.084008,Adami2020,Darabi_2022}.
Three-dimensional black hole solutions have been  widely  investigated from different physical viewpoints, to extend gauge field theory, the quantum theory of gravity,  and string
theory, in addition to studying the gravitational interaction in low dimensional manifolds \cite{Darabi2013,Carlip2005,SCarlip,cataldo2001btz,044027,sheykhi2021mimetic}.

It has been long known that perturbations of classical gravitational backgrounds described by black holes or branes naturally bring up the quasinormal modes (QNMs), which are damped oscillations with a discrete spectrum (see e.g. \cite{Konoplya,Nollert,Kokkotas2} for a detailed review).
The interest in studying the QNMs and their quasinormal frequencies (QNFs) dates back to the works of \cite{regge1957stability,zerilli1970effective,zerilli1970gravitational}.
Known also  as the ”ringing” of black holes,   the QNMs, their frequencies, and damping times are entirely determined by the black hole, being independent of the initial perturbation \cite{Hod2009}. 
Nowadays, QNMs  have particularly attracted interest since the observation of gravitational waves from the merger of two black holes \cite{Abbott}.
Although the observed signal shows consistency with the Einstein theory of gravity \cite{Abbott}, the uncertainties in the angular momenta and mass of the ringing black hole open the window for considering the alternative theories \cite{Gonzlez,konoplya2016detection}.
One of the applications of QNMs is studying the stability of matter fields, which without backreacting on the metric, evolve perturbatively outside the event horizon of the black holes \cite{Gonzlez}. 
Furthermore, in the context of AdS/CFT correspondence \cite{Maldacena1}, the QNMs provide information on how fast a thermal equilibrium can be reached in the boundary theory  \cite{Horowitz2}.
The connection between QNMs and Hawking radiation, studied for instance in \cite{CORDA}, is also one of the intriguing features of QNMs which has become increasingly important on the route to quantization of the gravity.

Being the solutions of the wave equation, the black hole QNMs are required to satisfy specific physical boundary conditions. At the event horizon only the pure ingoing waves are allowed by these boundary conditions \cite{Konoplya}. At the boundary at infinity, the boundary conditions may differ in various theories. For astrophysical purposes, pure outgoing waves are required, while in string theory the Dirichlet one is demanded \cite{Konoplya}.

Although the wave functions of QNMs are often described in terms of hypergeometric function \cite{Gonzlez,BTZQN}, the differential equations for the fermionic fields in some backgrounds, as it is the case in this paper, appear to be in the form of the  Heun’s equation.
The  Heun function is a generalization of the Gauss hypergeometric function and the associated equation is a second-order linear differential equation with four regular singular points \cite{ronveaux1995heun}. One of the differences between these two functions is the presence of an additional parameter called the accessory parameter in the Heun function. This parameter does not effect the characteristic exponents at any of the four singular points of the associated differential equation \cite{ronveaux1995heun}, and is known to play the role of eigen parameter in some physical applications of Heun function \cite{arscott1995heun}. Recently, this function has been encountered in gravity theories and astrophysics problems.
For instance, it is known that for the four-dimensional Kerr-Newman-de Sitter spacetime the Teukolsky equation can be transformed into Heun’s equation \cite{Suzuki_1998}, as well as the Klein-Gordon equation for a test scalar field for an asymptotically five-dimensional {AdS black holes} \cite{BarragnAmado_2019,Aliev_2009}. 
Some of the other applications of the Heun functions are in computing QNMs \cite{Hatsuda_2020,Oshita_2021,Novaes_2019,Kwon_2011,GURTASDOGAN2019134839}, wave scattering problems, and the Green’s function \cite{Motohashi_2021}, computing
the greybody factor \cite{Carneiro_da_Cunha_2016}, and Hawking radiation \cite{Gregory_2021,Nambu_2022,Oshita2022}.  Nevertheless, the Heun function has not been as widely used as the hypergeometric function in computing QNMs.  It is interesting to investigate the exact solution of fermionic perturbation 
in terms of the Heun function.

Among various three-dimensional black hole solutions, our main interest in this paper is on the BTZ black hole solution in string effective action \cite{Horowitz}, the  Lifshitz black hole solution in conformal gravity \cite{conformalbh}, and Lifshitz exact black hole solutions in Hu-Sawicki $F(R)$ theory with Hyperscaling violation \cite{hendi}. 
BTZ black hole solution, one of the well-known  $(2+1)$-dimensional black hole solutions,  was first obtained to Einstein's gravity model with negative cosmological constant in 
\cite{PhysRevLett.69.1849,PhysRevD.48.1506} and then modified as a solution to string theory in  \cite{Horowitz}.
It has long been known that the line element of the non-rotating BTZ is Weyl related to spacetime constructed by hyperbolic pseudosphere, which is constant negative Gaussian curvature Lobachevsky-type geometries. 
Here, we first establish the Weyl relation of BTZ black hole to a homogeneous $(2+1)$-dimensional spacetime, whose spatial part is a surface of negative constant Gaussian curvature that possesses the symmetries of two-dimensional Lie algebra and its line element is described by deformed hyperbolic function.
The homogeneous spacetimes,  known to have the symmetry of spatial homogeneity and constructed based on the simply-transitive  Lie groups classification \cite{Cosmictopology},  have been extensively used to construct cosmological and black hole solutions  \cite{Ellis1969,PhysRevD.57.5108,NADERI2017,Naderi2,bh}.
In addition to  BTZ black hole, exploring the previously known black hole solutions of three-dimensional gravity theories, we show that the Lifshitz black hole solution in three-dimensional conformal gravity obtained in \cite{conformalbh} and the Lifshitz exact solutions of Hu-Sawicki $F(R)$ model with Hyperscaling violation obtained in \cite{hendi}  share the properties that their line elements are Weyl-related to that of the mentioned homogeneous spacetime.

The Weyl relation of the line element of these three black hole solutions to that of the homogeneous spacetime enables us to take advantage of the local Weyl symmetry of the massless Dirac action to consider the matter distribution described by the fermionic field propagating outside the event horizon of these black holes by solving the same differential equations. In doing so,
deriving the Dirac equations and their solutions on the homogeneous spacetime, 
the wave function on the three black hole backgrounds can be obtained by employing proper Weyl transformations. 
The differential equations on the homogeneous spacetime are in the form of Heun’s equation. 
Using them, we compute QNM frequencies. 
Although the fermionic perturbation and QNMs of the two of the considered black holes, i.e. BTZ and the Lifshitz black hole solutions in conformal gravity, have been already studied in terms of   hypergeometric functions in \cite{Gonzlez,BTZQN,btzqnmhyper}, it is interesting to study these two backgrounds, a well as the Lifshitz {black} hole solutions of Hu-Sawicki $F(R)$ model,  in terms of the local Heun
function and its accessorize parameter. 

The solutions of the wave equations at the considered black hole backgrounds are obtained to be characterized by complex frequencies. 
Imposing the boundary conditions of QNMs at the event horizon and spatial infinity, we compute QNMs and quasinormal modes frequencies (QNFs) in terms of the black hole parameters and the accessory parameter of the Heun function.
Also, the stability of the fermionic field in these backgrounds has been studied by considering the exact QNFs.

The paper is organized as follows: 
Section \ref{se2} presents in detail the characteristics of the considered homogeneous spacetime and then lists the three black hole solutions, whose line element is Weyl related to that of the homogeneous spacetime.
Then, in section \ref{perturbations}, we solve analytically the Dirac equation in these backgrounds to find the  QNMs and study their stability. 
Finally, some concluding remarks are presented in section \ref{conclussion}. 

\section{$(2+1)$-dimensional black holes Weyl related to spatially homogeneous spacetime }\label{se2}

The $(2+1)$-dimensional spacetime, whose $t$-constant hypersurface is given by a homogeneous space corresponding to the $2$-dimensional Lie group with real two-dimensional Lie algebra $[T_1,T_2]=T_2$, can be described by the following The metric ansatz 
\begin{eqnarray}\label{metricc}
	ds^2=-dt^{2}+g_{ij}\sigma^{i}\sigma^{j},
\end{eqnarray}
where  $g_{ij}$ are constants and  left-invariant basis $1$-forms $\{\sigma^{i},~i=1,2\}$ {on the Lie group} obey $\sigma^{2}=-\frac{1}{2} 
\sigma^{1}\wedge \sigma^{2}$ and $\sigma^i=g^{-1}\partial^ig$, where $g={\rm e}^{x_1T_1} {\rm e}^{x_2T_2}$. The relations between coordinate and non-coordinate basis are given by
\begin{eqnarray}\label{sigma}
	\sigma^1=dx_{1}+
	x_2dx_2,\quad \sigma^2=dx_2.
\end{eqnarray}
Accordingly, the  metric \eqref{metricc} recasts the following form
\begin{eqnarray}\label{m1}
	\begin{split}
		ds^2=-dt^2&+(g_{11}+2 g_{12}\,x_2+g_{22} \,x_2^2)\,dx_1^2+2\left(g_{12}+g_{22}x_2\right)\,dx_1dx_2+g_{22}\,dx_2^2,
	\end{split}
\end{eqnarray}
whose Gaussian curvature is constant, given by
$$
{\cal{K}}=-\frac{g_{11}g_{22}-g_{12}^2}{g_{22}}.$$
Assuming ${\cal{K}}<0$ and defining a dimensionless parameter $k$ related to the amplitude of ${\cal{K}}$, i.e.
$k\equiv \mid{\cal{K}}\mid$,
the new coordinates $(\rho,\varphi)$ can be introduced via the   coordinate redefinition 
\begin{eqnarray}\label{}
	\begin{split}
		x_1&=-L\,   \varphi-\ln  \left(\frac{\left(2 k \,{\mathrm e}^{2 \rho  \sqrt{k}}+1\right) \sqrt{2}}{4 \sqrt{k}\, {\mathrm e}^{\rho  \sqrt{k}}}\right)
		,
		\\
		x_2&={\frac {1}{{4 g_{22}}\,k}}\left(\sqrt {{2\,g_{22}}} \left( 2\,{{\rm e}^{-\sqrt {k}   \rho
		}}k-{{\rm e}^{\sqrt {k}   \rho}} \right) -4\,{g_{12}}\,
		k\right)
		,
	\end{split}
\end{eqnarray}
where $L$ is a real constant. Then, the line element \eqref{m1} recast the following form
\begin{eqnarray}\label{m3}
	ds^2\equiv g_{\mu\nu}^Rdx^\mu dx^\nu=-dt^2+   \left(d\rho^2+R^2(\rho)d\varphi^2\right),
\end{eqnarray}
where
\begin{eqnarray}\label{m4}
	\begin{split}
		R(\rho)=q\left(2\,k\,{{\rm e}^{-\sqrt{k}   { \rho}}}+ {{\rm e}^{{
					{{ \sqrt{k} }    \rho}}}}\right)&=2q\cosh_{2k}(\sqrt{k}     \rho), \quad\quad {\rm with}\quad\quad q=\frac{\sqrt{2}L}{4\,k}.
	\end{split}
\end{eqnarray}
The $\cosh_{2k}$ function is known as the deformed hyperbolic function introduced for the first time in \cite{ARAI199163,EGRIFES2000229}, in solving  Schrodinger equation with deformed potential.

The line element \eqref{m3} is in the form of Lobachevsky-type metrics \cite{g}. Among the well-known Lobachevsky geometries, the Beltrami and elliptic type metrics can not be recovered from \eqref{m3}. But, setting the especial value of  $k=\frac{1}{2}$ can reduce the metric \eqref{m3} to a usual hyperbolic type. It was shown in \cite{cvetivc2012graphene} that the line element of BTZ black hole solution is related to the spacetime with hyperbolic space geometry by a Weyl transformation. However, in the following section, we show that the BTZ solution can be also described by a metric conformal related to the line element \eqref{m3}, described by a deformed hyperbolic function.

Furthermore, we explored some previously known black hole solutions in different theories of gravity to find the solutions Weyl related to the metric \eqref{m3}.  
In this regard, we also consider the two black hole solutions, including the Lifshitz black hole solution in three-dimensional conformal gravity obtained in \cite{conformalbh}, and the Lifshitz exact solutions of $F(R)$ obtained in \cite{hu2007models} and show their Weyl relation to the line element \eqref{m3}.

\subsection{BTZ Black hole solution of leading order string effective action }\label{se3}

BTZ solutions are black hole solutions to equations of motion of string effective action given by the $\beta$-function equations of $\sigma$-model, which are equivalent to the field equations of the associated gravity theory and assure the conformal invariance of the $\sigma$-model \cite{tseytlin1992elements}, given by the following line element \cite{Horowitz}
\begin{eqnarray}\label{mf}
	ds^2=-\left(\frac{{r}^{2}}{l^2}-m\right)
	dt^2+\left(\frac{{r}^{2}}{l^2}-m\right)^{-1}dr^2+ r^2d\varphi^2.
\end{eqnarray}
We will rewrite the AdS radius $l$ in terms of  the  central charge deficit of string theory $\Lambda$ by $l^2=\frac{2}{{\Lambda}}$, where   in non-critical $D$-dimensional bosonic theory $\Lambda$ is given by $	\Lambda=\frac{2\,(26-D)}{3\alpha'}$ \cite{tseytlin1992elements},  where the $\alpha'$ is square of string length,  $\alpha'=\lambda_s^2/2\pi$.

To rewrite \eqref{mf} in the form of line element Weyl related to \eqref{m3}, we apply the following redefinition
\begin{eqnarray}\label{ror}
	{ dr}={\frac {8\,km{{\rm e}^{\sqrt {\Lambda m}\rho}}}{
			\left( -2\,k+{{\rm e}^{\sqrt {\Lambda m}\rho}} \right) ^{2}}}{ d\rho},
\end{eqnarray}
which leads to $r=-{\frac {2\,\sqrt {m} \left( 2\,k+{{\rm e}^{\sqrt {\Lambda m} \rho}} \right) }{\sqrt {\Lambda} \left( -2\,k+{{\rm e}^{\sqrt {\Lambda m} \rho}} \right) }}
$. Now, if one sets
\begin{eqnarray}\label{m0}
	m=\frac{4 k    }{\Lambda},
\end{eqnarray}
then the BTZ metric \eqref{mf} recasts the following form
\begin{eqnarray}\label{m6}
	\begin{split}
		ds^2=&\frac{32    k^2}{\Lambda\left(2\,k\,{{\rm e}^{{{ -\sqrt{k} }    \rho}}}- {{\rm e}^{{	{{ \sqrt{k} }    \rho}}}}\right)^2}\bigg(-dt^2+d\rho^2+\frac{1}{2\Lambda k}\left(2\,k\,{{\rm e}^{{{ -\sqrt{k} }    \rho}}}+{{\rm e}^{{	{{ \sqrt{k} }    \rho}}}}\right)^2d\varphi^2\bigg).
	\end{split}
\end{eqnarray}
Since the metric \eqref{m6} has  diverging conformal factor at $\rho_0=\frac{\ln(2k)}{\sqrt{k}   }$, 
to have a well-defined coordinate redefinition we can restrict the range of $\rho$ coordinate to $\rho<\rho_0$, which is equivalent to $r>0$.
At the black hole event horizon $
r_{h}=\frac{4}{\Lambda} \sqrt{k}{ {{    } }{}}
$, we have $\rho\rightarrow -\infty$, while $r\rightarrow \infty$ corresponds to $\rho=\rho_0$.

\subsection{Lifshitz black hole solution in  three-dimensional conformal gravity}\label{conformalgr}

The Lifshitz  black hole solution for  three-dimensional conformal gravity has been obtained in \cite{conformalbh}, being described by the following line element
\begin{equation}\label{conformal}
	\begin{split}
		ds^2=-f(r)dt^2+\frac{l^2}{r^2}\frac{dr^2}{f(r)}+r^2d\varphi^2,\quad {\rm and}\quad f(r)=1-\frac{r_+^2}{r^2}.
	\end{split}
\end{equation}
It is an asymptotically Lifshitz black hole with dynamical exponent $z = 0$ and the event horizon located at $r = r_+$.
By a coordinate redefinition of the form 
\begin{equation}\label{r02}
	\begin{split}
		\frac{l}{r}\frac{dr}{f(r)}=d\rho,
	\end{split}
\end{equation}
which leads to $r=\sqrt {{{r_+}}^{2}+{{\rm e}^{{\frac {2(\rho+c)}{l}}}}}$, in which $c$ is an integrating constant, if one sets
\begin{equation}\label{e}
	\begin{split}
		c=\ln(q),\quad l^{-1}=\sqrt{k}   ,\quad r_+=\sqrt{2k}q,
	\end{split}
\end{equation}
the metric \eqref{conformal} takes the following form
\begin{equation}\label{mconfo}
	\begin{split}
		ds^2=&\frac{{\rm e}^{{{ \sqrt{k} }    \rho}}}{2\,k\,{{\rm e}^{{{ -\sqrt{k} }    \rho}}}+ {{\rm e}^{{
						{{ \sqrt{k} }    \rho}}}}}\bigg(-dt^2+d\rho^2+q^2\left(2\,k\,{{\rm e}^{{{ -\sqrt{k} }    \rho}}}+{{\rm e}^{{
					{{ \sqrt{k} }    \rho}}}}\right)^2d\varphi^2\bigg),
	\end{split}
\end{equation}
which shows that metric \eqref{conformal} is Weyl related to metric \eqref{m3} by a conformal factor that, despite that of the metric \eqref{m6},  has no diverging point to put a bound on the $\rho$ coordinate. Noting \eqref{r02}, at the black hole event horizon $
r_{+}$, we have $\rho\rightarrow -\infty$, while $r\rightarrow \infty$ corresponds to $\rho=+\infty$.

\subsection{ Lifshitz exact solutions of Hu-Sawicki $F(R)$ model with 	Hyperscaling violation} \label{Sawicki}

To find black hole solutions in  $F(R)$ gravity theories, some of the viable 
cases have been considered in the literature, in which local gravity
constraints are satisfied as well as cosmological and stability conditions. One of the viable $F(R)$ theories  is the so-called Hu-Sawicki model \cite{hu2007models}
\begin{equation}
	\label{fr}
	F(R)=R-m^{2}\frac{C_{1}\left( \frac{R}{
			m^{2}}\right) ^{n}}{1+C_{2}\left( \frac{R}{m^{2}}\right) ^{n}},
\end{equation}
for which the  asymptotically Lifshitz solution with a hyperscaling overall factor was obtained in \cite{hendi}, described by the following metric 
\begin{equation}
	ds^{2}=r^{\alpha }\left[ -\left( \frac{r^{2}}{l^{2}}\right) ^{z}f(r)dt^{2}+%
	\frac{l^{2}dr^{2}}{r^{2}f(r)}+r^{2}d\phi ^{2}\right] ,  \label{ALifMet}
\end{equation}%
where the constants $z$ and $\alpha $ denote the dynamical and hyperscaling violation exponents, respectively. In the particular case of  $\alpha =-2$, with the mentioned $F(R)$ model, the metric function {$f(r)$} is given by \cite{hendi}
\begin{equation}
	f(r)=\left( a+\frac{b}{r^{z-2}}\right) r^{-z}-\frac{l^{2}R_{0}}{%
		2r^{2}(z-2)^{2}}, \label{gLif}
\end{equation}%
where the constant $R_{0}$ is the Ricci scalar and 
the  solutions for  parameters of the model \eqref{fr} are
$
C_{1} =\frac{nm^{2n-2}}{R_{0}^{n-1}}$ and $
C_{2} =(n-1)\frac{m^{2n}}{R_{0}^{n}}$.

For this  black hole solution,
by a coordinate redefinition of the form 
\begin{equation}\label{}
	\begin{split}
		\frac{l^{1+z}}{r^{1+z}}\frac{dr}{f(r)}=d\rho,
	\end{split}
\end{equation}
which with setting $b=0$ leads to $$r= \left( {\frac {2\,a \left( z-2 \right) ^{2}{{\rm e}^{-{l}^{-1-
					z}a \left( z-2 \right)  \left( \rho+c_2 \right) }}+{l}^{2}{R_0}}{2a
		\left( z-2 \right) ^{2}}} \right) ^{- \left( z-2 \right) ^{-1}}
$$ in which $c_2$ is an integrating constant, if one sets
\begin{equation}\label{e1}
	\begin{split}
		&	c_2={\frac {1}{2\sqrt {k}    }\ln  \left( {\frac {{l}^{z-1}
					\left( z-2 \right) }{8{k}^{5/2}    \,{q}^{2}}} \right) }
		,\quad a=2\,{\frac {\sqrt {k}    \,{l}^{z+1}}{z-2}}, { R_0}=16\,{    }^{2}{k}^{2}{q}^{2},
	\end{split}
\end{equation}
where the radius of horizon becomes ${ r_h}= \left( {\frac {4{l}^{-z+1}
		{q}^{2}{k}^{\frac{3}{2}}    }{z-2}} \right) ^{- \left( z-2 \right) ^{-1}}
$,
the metric \eqref{ALifMet} takes the following form
\begin{equation}\label{mfrr}
	\begin{split}
		ds^2=&\frac{1}{q^2\left(2\,k\,{{\rm e}^{{{ -\sqrt{k} }    \rho}}}+ {{\rm e}^{{
						{{ \sqrt{k} }    \rho}}}}\right)^2}\big(-dt^2+d\rho^2+q^2\left(2\,k\,{{\rm e}^{{{ -\sqrt{k} }    \rho}}}+{{\rm e}^{{
					{{ \sqrt{k} }    \rho}}}}\right)^2d\varphi^2\big).
	\end{split}
\end{equation}
Similar to \eqref{mconfo}, the conformal factor in \eqref{mfrr} has no diverging point. At the black hole event horizon $
r_{h}$, we have $\rho\rightarrow -\infty$, while $r\rightarrow \infty$ corresponds to $\rho=+\infty$.

\section{Fermionic perturbations and Quasinormal modes}\label{perturbations}

We  consider the three aforementioned black hole solutions, whose metrics are conformally related to the metric of homogeneous spacetime \eqref{m3}, described by the general relation 
\begin{eqnarray}\label{m5}
	ds^2=h(\rho)g_{\mu\nu}^Rdx^\mu dx^\nu=h(\rho)\left(-dt^2+d\rho^2+R^2(\rho)d\varphi^2\right),\nonumber\\
\end{eqnarray} 
in which $R(\rho)$ is given by \eqref{m4} and the associated $h(\rho)$ functions for each of the considered black hole solutions have been determined in the previous section. 
Here, we consider matter distribution described by fermionic fields propagating outside the event horizon of these black holes. In this regard, it is convenient to take advantage of the local Weyl symmetry of the massless Dirac action under transformations that, in $(2+1)$ dimensional spacetime with metric \eqref{m5}, is described by
\begin{eqnarray}\label{conf}
	g_{\mu\nu}=h(\rho)g_{\mu\nu}^R  \quad {\rm and} \quad\Psi=h(\rho)^{-\frac{1}{2}}\Psi^R,
\end{eqnarray}
in which $\Psi$ and $\Psi^R$ are  two-component Dirac spinors propagating in the background described by $g_{\mu\nu}$  and $g_{\mu\nu}^R $, respectively. Finding the solutions for $\Psi^R$ on the homogeneous spacetime, the $\Psi$ can be obtained for each of the considered black hole backgrounds, using  the Weyl transformation \eqref{conf}.

\subsection{Fermionic perturbation on homogeneous spacetime}

In \eqref{conf},  $\Psi^R$ is a solution of the following Dirac equation
\begin{eqnarray}\label{dirac}
	\gamma^\mu{\cal{D}}_\mu\Psi^R=0,
\end{eqnarray}
in which, the curved spacetime $\gamma_\mu$ matrices are related to the flat spacetime  ${\hat{\gamma}}_a$ matrices by  vielbein $e_{\mu}^a$
\begin{equation}
	\gamma_\mu(x)=e_\mu^{~a}(x)\hat{\gamma}_a,
\end{equation}
where vielbein are defined by $g^R_{\mu\nu}(x) = e_\mu^{~a}(x)\;e_\nu^{~b}(x)\;\eta_{ab}$,  and $\eta_{ab}=\text{diag}(-1,1,1)$ is the  flat $2+1$ dimensional Minkowski metric. The ${\hat{\gamma}}^a$ matrices satisfy the standard Clifford algebra $\{{\hat{\gamma}}^a,{\hat{\gamma}}^b\}=2\eta^{ab}1$.
The covariant derivative $\cal{D}_\mu$ is defined respect to the metric $g_{\mu\nu}^R$ \eqref{m3} as follows 
\begin{equation}
	{\cal{D}}_{\mu}=\partial_\mu+\frac{1}{4}\,\omega_{\mu}^{ab}\,M_{ab},
	\label{eq:covder}
\end{equation}
where
$M_{ab}=\frac{1}{2}\,[{\hat{\gamma}}_a,{\hat{\gamma}}_b]$ are the Lorentz generators and  $\omega_{\mu}^{ab}$ in a torsion-free framework is given by  \cite{Eguchi:1980jx} 
\begin{equation}\label{spc}
	\omega_{\mu}^{~ab} = e_{\nu}^{~a}\partial_{\mu}e^{\nu b}+e_{\nu}^{~a}\Gamma_{\mu\lambda}^{~~\nu}e^{\lambda b},
\end{equation}
where $\Gamma_{\mu\lambda}^{~~\nu}$ stands for the affine connection.
Considering the metric \eqref{m3}, the vielbein for $g_{\mu\nu}^R$ are given as follows 
\begin{equation}
	e_0^{~0}=	e_1^{1}=1,\quad e_2^{~2}=
	R(\rho). 
\end{equation}	
Also, the non-zero components of the Levi-Civita connection for the metric are
\begin{equation}
	\Gamma^{1}_{22}=-R\,R'\;,\quad \Gamma^{1}_{12}=\Gamma^{1}_{21}=\ln(R)',
\end{equation}
which lead to the non-zero components of 
spin connection coefficients $\omega_{\mu}^{~ab}$ \eqref{spc} as follows
\begin{equation}\label{s}
	\omega_{2}^{~21}=-\,\omega_{2}^{~12}=-R'.
\end{equation}
Adopting the following curved spacetime gamma matrices representation choice   
\begin{equation}
	\gamma_\mu=e_\mu^{~a}\,\hat{\gamma}_a
	=\big(\, i \,\sigma_3\,,\;\sigma_1\,,\;R\:\sigma_2\,\big),
	\label{eq:gammacurv}
\end{equation}
where $\sigma_i$ are  Pauli matrices,  the Clifford algebra $\left\{\gamma^\mu,\gamma^\nu\right\}=2\,{g}^{R\,\mu\nu}\,$ is satisfied by $\gamma$ matrices.

A stationary state of the Dirac spinor is required to be single-valued at each point in spacetime. Hence, {$\Psi^R(t,\rho,\varphi)$} should be a periodic function in $\varphi$ with period $\varphi\in[0,2\pi]$. We consider {$\Psi^R(t,\rho,\varphi)$} in the following form
\begin{equation}\label{eq:Solution}
	\Psi^R(t,\rho,\varphi)={\rm e}^{-i\omega t}{\rm e}^{ik_{\varphi}\varphi}\frac{1}{\sqrt{R(\rho)}}\left(
	\begin{array}{cc}
		\psi_1^R(\rho) \\
		\psi_2^R(\rho)
	\end{array}
	\right),
\end{equation}
where $k_{\varphi}=0,\pm 1,\pm 2,...$  is the orbital angular momentum quantum number.  
Then, the Dirac equation \eqref{dirac} gives the coupled equations for the spinors as 
\begin{equation}\label{eq:19}
	\psi_1^{R}{'}-\frac{1}{R}k_\varphi
	\psi_1^R+\omega \psi_2^R=0,
\end{equation}
\begin{equation}\label{eq:20}
	\psi_2^{R}{'}+\frac{1}{R}k_\varphi\psi_2^R-\omega \psi_1^R=0.
\end{equation}
Combining the two equations, one can get the following decoupled equation for $\psi^R_1$ 
\begin{equation}\label{eq:EOMConstantFlux}
	\begin{split}
		&\psi_1^R{''}
		+\left(\frac{  k_\varphi R'-k_\varphi^2}{R^2}+\omega^2\right)\psi_1^R=0.
	\end{split}
\end{equation} 
Now,  substituting the $R(\rho)$ given by \eqref{m4} in \eqref{eq:19} and \eqref{eq:EOMConstantFlux}, and then defining the new variable $X(\rho)\equiv \exp(\sqrt{k}   \rho)$,  we obtain\footnote{We have chosen the new variable $X(\rho)$ particularity to facilitate the investigation the boundary conditions at the horizon and infinity in terms of the radial coordinate $\rho$. }
\begin{equation}\label{si2}
	\psi^R_{2}(X)=-\frac{\sqrt{k}   }{\omega}X\psi^R_{1}{}'(X)+\frac{k_{\varphi
		}X}{2\omega q(X^2+2k)}\psi^R_{1},
\end{equation} 
\begin{equation}\label{eq:EOMX1}
	\begin{split}
		&X^2\psi_1^R{''}(X) +X\psi_1^R{'}(X)
		+\frac{\omega^2}{k   } \psi_1^R(X)-\frac{2Xk_\varphi }{q   \sqrt{k}\left(2k+X^2\right)^2}
		\left(-X^2+2k+\frac{2k_\varphi}{q   \sqrt{k}}X\right) \psi_1^R(X)=0.
	\end{split}
\end{equation}  
These two equations will be solved to determine $\psi^R_1$ and $\psi^R_2$.

The equation \eqref{eq:EOMX1} has four regular singular points at $X=0,  \pm i\sqrt{2k},\infty$. 
Defining $Y\equiv \frac{iX}{\sqrt{2k}}$, where the singularity points are mapped into $Y=0,\mp 1,\infty$ and performing the following transformation 
\begin{equation}\label{psi2}
	\psi_1^R
	(Y)=(Y+1)^\xi (Y-1)^\nu Y^{\sigma}\chi_1(Y),
\end{equation} 
the equation \eqref{eq:EOMX1} can be rewritten as
\begin{equation}\label{eq:EOMX2}
	\begin{split}
		&\chi_1''(Y) +\frac{1}{Y(Y^2-1)}\left(2\, \left( \xi+\nu+\sigma+\frac{1}{2} \right) {Y}^{2}+2\, \left( \nu-
		\xi \right) Y-2\,\sigma-1
		\right)\chi_1'(Y)
		\\
		&+\frac{1}{2kq^2Y^2(Y^2-1)^2}\big[A\,Y^4+B\,Y^3+C\,Y^2+D \,Y+2\,q^2\left({k}{\sigma}^{2}+2\omega^{2}{ }^{-2}\right)\big]\chi_1(Y)=0,    	\end{split}
\end{equation}
in which 
\begin{equation}\label{eq:3d}
	\begin{split}
		A&=2kq^2\left( \left( \xi+\nu+\sigma \right) ^{2}+\frac{\omega^{2}}{   {k}}\right),
		\\
		B&={q}\left(2\sqrt {2}\,ik_{{\varphi}}k+2\,{ }^{2}q \left( 2(\nu+
		\sigma+\xi)-1 \right)  \left( \nu-\xi \right) {k}^{2}\right),
		\\
		C&=\frac{1}{k   }\big(-4\,k\omega^{2}q^2+4\,{k_{{\varphi}}}^{2}+2\,q^2{ } {k}^{2}\big( {\xi}^{2}+{\nu}^{2}-2 \left( \nu+\sigma+1
		\right) \xi-2 \left( \sigma+1\right) \nu-2\,{
			\sigma}^{2} \big) \big),\\
		D&=-2 q\left(- i k_{{\varphi}}\sqrt {2}+  kq\left( \nu-\xi
		\right)  \left( 2\,\sigma+1 \right)  \right).
	\end{split}
\end{equation}
The solution for this equation is of type   Heun functions. In such a way that  rearranging the parameters in terms of new parameters $  p,\alpha,\beta,\gamma$, and $\delta $  by assuming
\begin{equation}\label{30}
	\begin{split}
		\gamma&=2\sigma+1,\quad\quad \sigma^2=\frac{\omega^2}{k   },\\
		&2\,\nu+2\,\xi+2\,\sigma-\beta-\alpha=0,
		\\
		&2\,\nu-2\,\xi- \left( -\delta-\gamma \right) \alpha+\alpha-\delta+\beta+1=0,\\
		&\alpha\beta-\frac{A}{2kq^2}=0,\quad\quad -\alpha\beta-\frac{C}{2kq^2}=0,\\
		&-p-\frac{B}{2kq^2}
		=0,\quad\quad p-\frac{D}{2kq^2}
		=0,
	\end{split}
\end{equation}  
the equation \eqref{eq:3d} takes the following form \cite{ronveaux1995heun}
\begin{equation}\label{kk}
	\begin{split}
		\chi_1'' \left( Y \right) &+\frac {
			\chi_1' \left( Y \right) }{Y \left( {Y}^{2}-1 \right) } \bigg( \left( \alpha+\beta+1 \right) {Y}^{2}+ \left(  \left( -\delta-\gamma
		\right) \alpha-\alpha+\delta-\beta-1 \right) Y-\gamma \bigg)+{\frac { \left( \alpha\beta Y-
				p \right)  }{Y \left( {Y}^{2}-1 \right) }}\chi_1 \left( Y \right)=0.
	\end{split}
\end{equation}

Before solving the equations, it is worth providing the forms of Heun equation parameters in terms of the black hole solution parameters.
According to \eqref{30}, the powers in \eqref{psi2} take the following forms
\begin{equation}\label{alfabeta}
	\begin{split}
		&	\xi=\frac{1}{2q}\left(q(1-\kappa_1)-{\frac {i\kappa_1\,k_{{\varphi}}\sqrt {2}}{k }}\right), 
		\quad\qquad \nu=\frac{1}{2q}\left(q(1-\kappa_2)+{\frac {i\kappa_2\,k_{{\varphi}}\sqrt {2}}{k }}\right),
	\end{split}
\end{equation}
where $\;\;\kappa_{1},\kappa_{2}=\pm 1$,  and
\begin{equation}\label{sigmaa}
	\begin{split}
		\sigma={\frac {i\eta \,w}{\sqrt {k} }},\quad {\rm where }\;\; \eta =\pm 1.
	\end{split}
\end{equation}
Also,  \eqref{30} leads to
\begin{equation}\label{constants}
	\begin{split}
		\alpha&=\xi+\nu +{\frac {2\,i\eta \,w}{\sqrt {k} }},\quad \beta=\xi +\nu , \quad \gamma={\frac {2\,i\eta \,w}{\sqrt {k} }}+1,\\
		\delta&=\left( {\frac {2\,i\eta \,w}{\sqrt {k}}}+\xi+\nu -1 \right) ^{-1}\left( {\frac {4\omega^{2}}{k{ }^{2}}}-{
			\frac {2\,i \left( \nu +\xi +2 \right) w\eta }{\sqrt {k} }}-5
		\,\nu -\xi -1 \right)  .
	\end{split}
\end{equation}

Also,
the necessary conditions  \eqref{30}, for the  solution, to be a polynomial of the general Heun functions gives the energy levels as follows
\begin{equation}\label{eq:31}
	\omega={\frac {-k_{{\varphi}}\sqrt {2}+iqk  \left( 2\,{\xi }^{2
			}-2\,{\nu }^{2}-p -\xi +\nu  \right) }{2\eta\,q\sqrt {k} \, \left( 
			\xi  -\nu\right) }},
\end{equation}
which, using \eqref{alfabeta}, will be rewritten in terms of the parameters $k$, $k_\varphi$, $q$, and $p$ for each of black hole solutions.
Known as the a accessory or auxiliary parameter, 
$p$ is a complex number that has no impact on the singular points of the equation. 
In some physical applications of the Heun function, this parameter can play the role of eigen parameter \cite{arscott1995heun}.

According to \eqref{eq:31},  the solutions possess complex frequency, which is known to be the characteristic of  QNMs.
The QNMs are required to satisfy specific boundary conditions at the
black hole event horizon as well as the infinity. At the event horizon, the condition is the existence of only ingoing waves and at the infinity, the outgoing waves or Dirichlet boundary conditions are required \cite{Konoplya}. 
Hence, we are interested in finding solutions to equation \eqref{kk}  around the event horizon and far from the horizon.

As said before, the equation \eqref{eq:EOMX1} has four regular singular points at $X=0,\infty,  \pm i\sqrt{2k}$. One of the interesting points is $X=0$, which is equivalent to $\rho\rightarrow -\infty$ that happens at the event horizon for all of the considered three types of black holes.
The $X=\infty$, on the other hand, is equivalent to $\rho=\infty$ and $r\rightarrow \infty$ for the  Lifshitz black hole solution in conformal gravity, provided in section \ref{conformalgr}, and Lifshitz black hole solutions of Hu-Sawicki model, provided in section \ref{Sawicki}. 
Despite these two cases, for which the range of $\rho$ is complete, i,e, $\rho\in(-\infty,+\infty)$,  for black hole solution of string effective action $\rho$ ranges in $-\infty<\rho<\rho_0$. For this class of solutions, at $\rho=\rho_0$ $r\rightarrow \infty$ and $X=\sqrt{2k}$, which is not one of the singular points of the equation \eqref{eq:EOMX1}. Consequently, although the form of the differential equations of $\Psi^R$ are the same for the three backgrounds, the solution at the boundary far from the black hole event horizon should be obtained in different ways.

We will provide the solutions at both boundaries for the considered black hole solutions in the following subsections. {It should be noted that the Heun function parameters in \eqref{alfabeta}-\eqref{constants} do not satisfy the special relations that allow the transformation of Heun functions to hypergeometric functions.  Hence, we will analyze the solutions in terms of the Heun functions.}

Before continuing, it is worth preparing the general formula for energy flux for the considered black holes as well as the stability analysis formula to be used on the solutions.

\subsubsection{Energy flux Formula}
The energy-momentum tensor for the Dirac field is defined by 
\begin{equation}\label{}
	\begin{split}
		T_{\mu\nu}=\frac{i}{8\pi}\Psi^{\dagger}\gamma^0\left[\tilde{\gamma}_\mu\tilde{{\cal D}}_\nu+\tilde{\gamma}_\nu\tilde{{\cal D}}_\nu\right]\Psi+cc,
	\end{split}
\end{equation}
where c.c. indicates the complex conjugate
of the preceding terms. Also, $\tilde{\gamma}_\mu$ and $\tilde{{\cal D}}_\mu$ are the gamma matrix and the covariant derivative associated with  the metric $g_{\mu\nu}$
and $\tilde{\gamma}_\mu=\sqrt{h(\rho)}\gamma_\mu$, where $\gamma_\mu$ are given by \eqref{eq:gammacurv}. The
energy flux at radial coordinate $\rho$ is defined by 
\begin{equation}\label{}
	\begin{split}
		{\cal F}\mid_\rho=\int \sqrt{-g}d\varphi T^\rho_{\,t},
	\end{split}
\end{equation}
Noting \eqref{s}, we have $\tilde{\omega}_0^{ab}=\tilde{\omega}_\rho^{ab}=0$.

For the considered black holes, which admit the  conformal transformation  \eqref{conf} in which the Dirac spinors are considered in the form of \eqref{eq:Solution}, we get
$$ T^\rho_{\,t}=\frac{i(\omega+\omega^*)}{4\pi R\sqrt{h}}\left(\psi_1^R\psi_2^{R*}-\psi_2^R\psi_1^{R*}\right),$$ where $\omega^*$ is the complex conjugate of $\omega$. Consequently, the energy flux takes the form
\begin{equation}\label{F}
	\begin{split}
		{\cal F}\mid_\rho\propto i\left(\psi_1^R\psi_2^{R*}-\psi_2^R\psi_1^{R*}\right).
	\end{split}
\end{equation}
Interestingly, the energy flux appeared to be independent of the $h(\rho)$ function in \eqref{conf}. 

\subsubsection{Stability analysis }

Reminding that the time-dependent part of the wave function \eqref{eq:Solution} is ${\rm e}^{-i\omega t}$, writing QNM frequencies in the form of $\omega=\omega_{Re}+i \omega_{Im}$, positive values of the imaginary part of $\omega$ means instability, while the negative valued ones mean that $\Psi$ is damped \cite{Konoplya}. In fact,  $\omega_{Re}$ is the real oscillation frequency of the mode, and the $\omega_{Im}$ is proportional to the damping rate of mode \cite{Konoplya}. Nevertheless, it is worth analyzing the stability of the solutions following the procedure presented in \cite{Gonzlez,Ortega}.

Considering the equation \eqref{eq:EOMConstantFlux} in the notion of $\psi_1^R{''}
+\left(-V_1(\rho)+\omega^2\right)\psi_1^R=0$,   the potential $V_1(\rho)=-\frac{  k_\varphi R'-k_\varphi^2}{R^2}$, which recasts the following form by substituting the  $R(\rho)$ given by \eqref{m4} 
\begin{equation}\label{V}
	V_1(\rho)=\frac {2\,k_{{\varphi}}{{\rm e}^{\rho\,\sqrt {k}   }}\sqrt {k}   }{{q}^{2
		} \left( {{\rm e}^{2\,\rho\,\sqrt {k}   }}+2\,k \right) ^{2}} \left( {
		{\rm e}^{2\,\rho\,\sqrt {k}   }}q+{\frac {2\,k_{{\varphi}}{{\rm e}^{\rho
					\,\sqrt {k}   }}}{\sqrt {k}   }}-2\,qk \right) .
\end{equation}
Now,  multiplying equation \eqref{eq:EOMConstantFlux} by {$\psi_1^{R*}$} and  integrating by parts, one can arrive at the following expression
\begin{equation}\label{stab}
	\begin{split}
		\int_{\rho_h}^{\infty}\left(\mid \psi_1^R{'}\mid^2+V_1(\rho)\mid \psi_1^R\mid^2\right)d\rho=&\omega^2\int_{\rho_h}^{\infty}\mid \psi_1^R\mid^2d\rho+k_{\varphi}\left(
		\frac{\mid \psi_1^R\mid^2}{R(\rho)}\mid_{\rho=\infty}-\frac{\mid \psi_1^R\mid^2}{R(\rho)}\mid_{\rho=\rho_h}\right)\\
		&+\omega\left(\psi_1^R {}^{*}\psi_2^R\mid_{\rho=\rho_h}-\psi_1^R {}^{*}\psi_2^R\mid_{\rho=\infty}\right),
	\end{split}
\end{equation}
where for BTZ black hole solution  $\rho_h=\rho_0$ and for the other considered black holes $\rho_h=-\infty$. It is worth noting that the $V_1(\rho)$ in \eqref{V} is not positive-definite.
Also, $R(\rho)^{-1}$ vanishes for $\rho=\pm \infty$ and equals to ${2k}^{-\frac{1}{2}}$ for $\rho=\rho_0$. 

Now, following the deformation method provided in \cite{Gonzlez,Ortega}, defining the new derivative $D=\frac{d}{d\rho}+S(\rho)$, the integral in the right hand of equation \eqref{stab} becomes
\begin{equation}
	\begin{split}
		&	\int_{\rho_h}^{\infty}\left(\mid \psi_1^R{'}\mid^2+V_1(\rho)\mid \psi_1^R\mid^2\right)d\rho=\int_{\rho_h}^{\infty}\left(\mid D\psi_1^R\mid^2+\tilde{V}_1(\rho)\mid \psi_1^R\mid^2\right)d\rho-S(\rho)\mid \psi_1^R\mid^2\mid_{\rho=\rho_h}^{\rho=\infty},
	\end{split}
\end{equation}
where $\tilde{V}_1(\rho)={V}_1(\rho)+S'-S^2$. 
To ensure the stability of the spacetime against the perturbation field,  a  $S(\rho)$ function needs to be found  such that the new potential  satisfies
$\tilde{V}_1\geq 0$ \cite{Gonzlez,Ortega}. 
The appropriate function here is $S(\rho)=-\frac{k_{\varphi}}{R(\rho)}$, which leads to $\tilde{V}_1(\rho)=0$. Then, \eqref{stab} recasts the following form
\begin{equation}\label{stab2}
	\begin{split}
		\int_{\rho_h}^{\infty}\mid D\psi_1^R\mid^2&d\rho=\omega^2\int_{\rho_h}^{\infty}\mid \psi_1^R\mid^2d\rho+\omega\left(\psi_1^R {}^{*}\psi_2^R\mid_{\rho=\rho_h}-\psi_1^R {}^{*}\psi_2^R\mid_{\rho=\infty}\right).
	\end{split}
\end{equation}
The integral is  positive. The remaining step is to consider the sign of the last terms on the right hand of the equation \eqref{stab} using the asymptotic behavior of the obtained solutions at the boundaries,  which will be studied in the following sections.\footnote{If the equation for $\psi^R_2$ is considered, which is obtained by combining \eqref{eq:19} and \eqref{eq:20} as $\psi_2^R{''}
	+\left(-V_2(\rho)+\omega^2\right)\psi_2^R=0$,  where the potential $V_2(\rho)=\frac{  k_\varphi R'+k_\varphi^2}{R^2}$, the same procedure by choosing $S(\rho)=\frac{k_{\varphi}}{R(\rho)}$ can be applied.}

Having prepared the differential equations to be solved as well as the form of energy flux, we continue to find the wave functions to compute the QNMs and impose the boundary conditions and stability conditions.

\subsection{Dirac QNMs of BTZ black hole solution  in terms of Heun function}

To obtain the Dirac spinor $\Psi$ defined by \eqref{conf}, which propagates outside the event horizon of the BTZ black hole \eqref{mf}, we need to find $\psi_1^R$ and $\psi_2^R$ in \eqref{eq:Solution}. In doing so, we first solve the equation \eqref{kk} and  use \eqref{psi2} to determine $\psi_1^R$. Substituting   $\psi_1^R$ into \eqref{si2} will then give the $\psi_2^R$. 
The constant $q$ in \eqref{m4} for this class of solutions  has been fixed as $q^2={\frac {1}{2 {\Lambda}k}}$. Also, the associated  $h(\rho)$ function in \eqref{conf} can be reed of \eqref{m6}.
As mentioned before, for this black hole solution the region $r_h \leq r < \infty$ of our interest amounts to the non-complete range $-\infty < \rho \leq \rho_0=\frac{\ln(2k)}{\sqrt{k}   }$, where the $\rho$ is defined in terms of $r$ by \eqref{ror}. We are particularly interested in finding solutions at the black hole event horizon and spatial infinity.

The event horizon of the black hole, in terms of $X(\rho)=\exp(\sqrt{k}   \rho)$ function, is mapped into $X=0$, or equivalently $Y=0$, which is one of the regular singular points of the equation  \eqref{kk}. The solution at the event horizon, denoted by $\chi_{i0}$, is given by the  following combination of local Heun functions $Hl$
\begin{equation}\label{xi1}
	\begin{split}
		\chi_{10}&(Y)=C_{1}{\it Hl} \left( -1,p ,\alpha ,\beta ,\gamma,\delta ,Y \right)+C_{2} Y^{1-\gamma}{\it Hl} \left( -1,p_2 ,\alpha +1-\gamma,\beta +1-\gamma,-\gamma+
		2,\delta,Y \right)
		,
	\end{split}
\end{equation}
where $C_{1}$, $C_{2}$ are constant and $$p_2=p + \left( \gamma-1 \right)  \left( 2\,\delta +
\gamma-\alpha -\beta -1 \right).$$ The $\alpha ,\beta ,\gamma,$ and $\delta $ are  complex parameters defined by \eqref{alfabeta}-\eqref{constants}.

On the other hand, at $\rho=\rho_0$, which corresponds to the radius far from the horizon,  we have   $Y=i$ that is not one of the singular points of the equation \eqref{kk}. To find the solutions at this limit, we use the linear transformation $Y\rightarrow -Y+i$, which moves the singular point of $Y=0$ to $Y=i$,  where the solution for \eqref{kk} is given in the following form
\begin{equation}\label{xi11}
	\begin{split}
		\chi_{11}(Y)&=C_3{\it Hl} \left( -1,p ,\alpha ,\beta ,\gamma,\delta ,i-Y \right)+C_4 (i-Y)^{1-\gamma}{\it Hl} \big( -1,p_2 ,\beta +1-\gamma,\alpha +1-\gamma,-\gamma+
		2,\delta,i-Y\big)
		,
	\end{split}
\end{equation}
where $C_{3}$, $C_{4}$ are constant. 

Then, using \eqref{psi2}, the solutions for $\psi_1^R$ near the horizon and at the radius far from the horizon  are given, respectively, by 
\begin{equation}\label{Psi1}
	\psi^R_{10}(X)=\left(\sqrt{2k}+iX\right)^{\xi } \left(\sqrt{2k}-iX\right)^{\nu } X^\sigma \chi_{10},
\end{equation}
\begin{equation}\label{Psi1in}
	\begin{split}
		\psi^R_{11}(X)=\left(\sqrt{2k}-i(X-\sqrt{2k})\right)^{\xi } &\left(\sqrt{2k}+i(X-\sqrt{2k})\right)^{\nu } (\sqrt{2k}-X)^{\sigma }\chi_{11}.
	\end{split}
\end{equation}

Furthermore, using \eqref{Psi1} and \eqref{Psi1in}, equation \eqref{si2} yields
\begin{equation}\label{psi22}
	\begin{split}
		\psi^R_{2}&(X)=\frac{  1 }{q\omega}\left(\sqrt{2k}+ix\right)^{\xi } \left(\sqrt{2k}-ix\right)^{\nu } x^{\sigma }\bigg[\frac{i \sqrt {2}q}{2}C_{13}x\,{\it Hlp} \left( -1,p,\alpha,\beta,\gamma,\delta,
		\frac{ix}{\sqrt{2k}}\right)\\
		&-\frac {\sqrt {k}C_{13}{ Hl} \left( -1,p,\alpha,\beta,\gamma,\delta,\frac{ix}{\sqrt{2k}}
			\right)}{{x}^{2}+2\,k} \big( -2\,qk\sigma+x \left( -x
		\left( \xi+\nu+\sigma \right) q+{\frac {2k_{{\varphi}}}{{	\sqrt {k}    }
		}}+i\sqrt {2k} \left( \nu-\xi \right) q \right) 
		\big)\\
		& -C_{24}x^{1-\gamma}\bigg(\frac {\sqrt {k} }{{x}^{2}+2k} {Hl} \left( -1,{p_2},\alpha_2,\beta_2,2-
		\gamma,\delta,\frac{ix}{\sqrt{2k}} \right)
			\big( -2q k\left( \sigma+
		1-\gamma \right) +x ( -x\left( \xi+\nu-\gamma+\sigma+1 \right) 
		q\\
		&+{\frac {2k_{{\varphi}}}{{   }\,\sqrt {k}}}+iq\sqrt {2k} \left( 
		\nu-\xi \right)  )  \big)-\frac{i \sqrt {2}\,q}{2}x{Hlp} \left( -1,p_2,\alpha_2,\beta_2,2-\gamma,
		\delta,\frac{ix}{\sqrt{2k}} \right)
		\bigg)
		\bigg],
	\end{split}
\end{equation}
in which for the solutions near the horizon, namely the $\psi^R_{20}(X)$, we set $x=X$, $C_{13}=C_1$, and $C_{24}=C_2$, while  for the solutions at spatial infinity, namely the $\psi^R_{21}(X)$, we set $x=-X+\sqrt{2k}$,  $C_{13}=\sqrt{2}C_3$, and $C_{24}=\sqrt{2}C_4$. Also, $\alpha_2=\alpha+1-\gamma$ and $\beta_2=\beta+1-\gamma$.

Accordingly, at the event horizon, where $\rho\rightarrow -\infty$ and  $X=0$, using the parameters given by \eqref{sigmaa} and \eqref{constants},
the $\psi^R$ components behave as 
\begin{equation}\label{horizon}
	\begin{split}
		&	\psi_{1h}^R=(-1)^\nu(2k)^{\frac{\xi+\nu}{2}}\left( C_{1} {\rm e}^{{\frac{i\eta\omega}{\sqrt{k}   }}\rho}+C_{2} \,{\rm e}^{-\frac{i\eta\omega}{\sqrt{k}   }\rho}\right),
		\\& 
		\psi_{2h}^R =i\sqrt{2}(-1)^\nu(2k)^{\frac{\xi+\nu}{2}}\left(   C_{1} {\rm e}^{{\frac{i\eta\omega}{\sqrt{k}   }}\rho}-C_{2} \,{\rm e}^{-{\frac{i\eta\omega}{\sqrt{k}   }}{}\rho}\right),
	\end{split}
\end{equation} 
in which the identities
$
{\it Hl} \left( -1,p,\alpha,\beta,\gamma,\delta,0\right)=1
$ and $\,$ ${\it Hlp} \left( -1,p,\alpha,\beta,\gamma,\delta,0\right)=-\frac{p}{\gamma}$ have been used. With $\eta=1$ ($\eta=-1$) the  second (first) terms in $\psi_{1h}^R$ and $\psi_{2h}^R$ describe ingoing waves.

Also, at $r\rightarrow \infty$,  where $X=\sqrt{2k}$ and $Y=i$,   the $\psi^R$ behave as 
\begin{equation}\label{asymp1}
	\begin{split}
		\psi_{1\infty}^R(X)=& \left( -1 \right) ^{\nu} \left( -\sqrt {2k} \right) ^{\frac{i\eta\omega}{\sqrt{k}   }+\xi+\nu}\bigg(
		C_{3}\,{\rm e}^{\frac{i\eta\omega}{\sqrt{k}   }\ln(1-\frac{X}{\sqrt{2k}})}+C_{4}\,\left( -\sqrt {2k} \right) ^{-\frac{2i\eta\omega}{\sqrt{k}   }} \,
		{\rm e}^{-\frac{i\eta\omega}{\sqrt{k}   }\ln(1-\frac{X}{\sqrt{2k}})}\bigg)
		,\\
		\psi_{2\infty}^R(X)=&i \left( -1 \right) ^{\nu} \left( -\sqrt {2k} \right) ^{\frac{i\eta\omega}{\sqrt{k}   }+\xi+\nu}\bigg(-
		C_{3}\,{\rm e}^{\frac{i\eta\omega}{\sqrt{k}   }\ln(1-\frac{X}{\sqrt{2k}})}+C_{4}\,\left( -\sqrt {2k} \right) ^{-\frac{2i\eta\omega}{\sqrt{k}   }} \,
		{\rm e}^{-\frac{i\eta\omega}{\sqrt{k}   }\ln(1-\frac{X}{\sqrt{2k}})}\bigg).
	\end{split}
\end{equation}
Noting $\ln(1-\frac{X}{\sqrt{2k}})<0$, with $\eta=-1$ ($\eta=1$) the  first (second) terms  describe  outgoing waves in $\psi^R$.  

We are interested in considering the boundary conditions on the $\Psi(t,X,\varphi)$ defined by \eqref{conf}, as the solutions of the wave equation for the BTZ black hole background,  for which  the  $h(\rho)$ function in \eqref{conf} can be read from the metric \eqref{m6}. Having found the components of $\Psi^R$ and their behaviors at the boundaries, the $\Psi$ can now be determined noting  \eqref{conf}  and \eqref{eq:Solution}
as follows
\begin{equation} \label{d}
	\Psi(t,X,\varphi)=\frac{1}{4lk\sqrt{q}}\frac{X^2-2k}{\sqrt{X(X^2+2k)}}{\rm e}^{-i\omega t}{\rm e}^{ik_{\varphi}\varphi}\left(
	\begin{array}{cc}
		\psi_1^R(X) \\
		\psi_2^R(X)
	\end{array}
	\right).
\end{equation}

At the horizon,
using \eqref{horizon}, the energy flux \eqref{F}  recasts the following form
\begin{equation}\label{fh}
	{\cal F}_h\propto-
	\eta\, \left( C_1C_1^*-C_2C_2^*  \right) ,
\end{equation}
As we expected from \eqref{horizon}, on the event horizon, with $\eta=-1$ ($\eta=1$) the second (first) term in \eqref{fh} represents an outgoing flux, which is not allowed on this boundary.  
Hence, noting \eqref{horizon} and \eqref{d}, to have pure ingoing waves and energy flux at the event horizon,  the $\eta=1$ and $\eta=-1$ cases should be accompanied by $C_{1}=0$ and $C_{2}=0$, respectively.

Considering the equation \eqref{V}, the potential $V_1$ vanishes at the event horizon $r_h$, where $\rho\rightarrow -\infty$, while at $r\rightarrow \infty$ where $\rho=\rho_0$,  we have $V_1=\frac{4}{l^2} k_\varphi^2$. Hence, despite the notion of the perturbed BTZ black hole with the electromagnetic field in \cite{BTZQN}, whose potential diverges at infinity, the equation  \eqref{eq:EOMConstantFlux} does not itself prescribe imposing the Dirichlet boundary condition with vanishing wave functions at infinity.
However, as mentioned before the range of $\rho$ coordinate is not complete in the considered class of black hole solution \eqref{mf}.
In the cases that the coordinate has an upper bound, similar to that of the asymptotically anti-de Sitter spacetimes \cite{Konoplya}, the simplest case of $\Psi\rightarrow 0$ at $r\rightarrow \infty$ is usually adopted \cite{Kodama_2004}.
Here, noting \eqref{asymp1} and \eqref{d}, the wave function $\Psi$ component are proportional to $C_3(X-\sqrt{2k})^{1-\eta\frac{\omega_{Im}}{\sqrt{k}   }}$ and $C_4(X-\sqrt{2k})^{1+\eta\frac{\omega_{Im}}{\sqrt{k}   }}$ at distances far from black hole. Reminding the $\omega_{Im}<0$ condition required for stability of QNMs,  for the $\eta=1$ and $\eta=-1$ cases the wave function terms, respectively, associated with $C_3$ and $C_4$ vanish at the infinity defined by $X=\sqrt{2k}$, satisfying the Dirichlet boundary condition. This is consistent with the specific boundary conditions of string theory solutions, which require vanishing wave functions at infinity \cite{Konoplya}.

Using \eqref{asymp1},  the energy flux \eqref{F}  at infinity becomes
\begin{equation}\label{fasy1}
	{\cal F}_{asymp}\propto
	-\eta\, \left( C_3C_3^*-C_4C_4^* \right) .
\end{equation}
Hence, the $C_1=0$ and $C_4=0$ ($C_2=0$ and $C_3=0$) conditions for the $\eta=1$ ($\eta=-1$) are equivalent to accepting only the ingoing and outgoing energy flux, respectively, at the horizon and spatial infinity.
Considering these conditions, and using the asymptotic behavior of the solutions given by \eqref{horizon} and \eqref{asymp1} in the last terms of the equation \eqref{stab2},  it can be concluded that the negative imaginary part of $\omega$ is consistent with the positivity of the integral in\eqref{stab2},  guaranteeing the stability of the fermionic perturbation outside of the black hole event horizon.

Finally, using \eqref{m0} and \eqref{alfabeta} in  \eqref{eq:31}, the corresponding energy levels associated with the different combinations of $\kappa_{1}$ and $\kappa_{2}$  are given  by
\begin{equation}\label{w3}
	\omega 
	=\frac{2}{\eta l}\left(- k_{{\varphi}}+\frac{1}{4l}ir_h\left(1-p \right)\right)
	\quad{\rm with}\quad \kappa_{1}=-\kappa_{2}=-1,
\end{equation}
\begin{equation}\label{w4}
	\omega 
	=\frac{2}{\eta l}\left( k_{{\varphi}}+\frac{1}{4l}ir_h\left(1+p \right)\right)
	\quad{\rm with}\quad \kappa_{1}=-\kappa_{2}=1,
\end{equation}
\begin{equation}\label{w2}
	\omega 
	=\frac{1}{l^2\eta}\left(- {\frac {r_h^2}{8l k_{
				{\varphi}}}}\,p+i r_h\right) 
	\quad{\rm with}\quad \kappa_{1}=\kappa_{2}=-1,
\end{equation}
\begin{equation}\label{w1}
	\omega 
	={\frac {r_h^2 }{8 l^3 \eta k_{{
					\varphi}}}} p
	\quad
	{\rm with}\quad \kappa_{1}=\kappa_{2}=1,
\end{equation}
which are expressed in terms of the 
accessory parameter $p$, which is a characteristic of the Heun function.
With the complex-valued  $p$, i.e. $p=p_{Re}+ip_{Im}$,
the imaginary part of $\omega$ is labeled by for \eqref{w1} and \eqref{w2} by $p_{Im}$, while for \eqref{w3} and \eqref{w4} it is labeled by $p_{Re}$.
The $\omega_{Im}$ grows monotonically with $p$. 
 The stability of the QNMs affects the choices of $p$ as the negative imaginary part of QNFs requires $p_{Re}>1$ ($p_{Re}<1$) for $\eta=1$ ($\eta=-1$) in \eqref{w3}, $p_{Re}<-1$ ($p_{Re}>-1$) for $\eta=1$ ($\eta=-1$) in \eqref{w4}, $\frac{\mathit{p_{Im}}}{k_{\varphi}}>\frac{8 l}{\mathit{rh}}$ ($\frac{\mathit{p_{Im}}}{k_{\varphi}}<\frac{8 l}{\mathit{rh}}$) for $\eta=1$ ($\eta=-1$) in \eqref{w2}, and  $\frac{{p_{Im}}}{k_{\varphi}}<0$ ($\frac{{p_{Im}}}{k_{\varphi}}>0$) for $\eta=1$ ($\eta=-1$) in \eqref{w1}.

Considering the dimensional analysis of the metric \eqref{mf},   $l$ is a constant
with dimensions of length \cite{Horowitz}, i.e. $\sqrt{\Lambda}$ has dimension of inverse length, 
$\varphi$ is dimensionless and hence the $k_\varphi$ in the wavefunction \eqref{eq:Solution} is dimensionless. Also, $\eta$ is chosen to be dimensionless parameters. Hence, the obtained   QNFs \eqref{w3}-\eqref{w1} are in the proper dimension of inverse length.

Using the hypergeometric function, the  QNMs and associated frequencies of BTZ black hole have been calculated in \cite{btzqnmhyper} in the form of $\omega=\pm \frac{k_{\varphi}}{l}+i\frac{4r_h}{l^2}(n+1)$, with $n=0,1,2,\dots$ as the overtone number. Here, by choosing  $p =\pm(16n+15)$ and noting that the $Hl(-1, \pm p, \alpha, \beta, \gamma, \delta, Y)$ functions are both solutions of the differential equation \eqref{kk} \cite{abramowitz1968handbook}, the QNFs in the cases of \eqref{w3} and \eqref{w4} can recast a similar form to those of \cite{btzqnmhyper}.\footnote{It is worth mentioning that QNFs for BTZ  have been also found by \cite{BTZQN} via hypergeometric function, leading to $w=\pm k_\varphi-\frac{2ir_h}{l}(n+1)$. Although it is not in the proper dimension of inverse length, by choosing  $p =\pm(2n+1)$ the QNFs in the cases of \eqref{w3} and \eqref{w4} can recast a similar form.}
In addition, using the Heun function the QNFs have been calculated for BTZ black hole in \cite{Kwon_2011} in the form of $\omega=\pm k_{\varphi}+2ir_h(n+1)$. Although they were introduced in the dimension of length, they resemble the  QNFs  \eqref{w3} and \eqref{w4}. 


Although the QNFs \eqref{w3} and \eqref{w4} can be reduced to those obtained in the procedures based on the hypergeometric function \cite{btzqnmhyper} and Heun function \cite{Kwon_2011},  the  QNFs \eqref{w1} and \eqref{w2} are particularly introduced in our formalism based on the Heun function. 
It is also interesting to note that the QNFs of \eqref{w1} and \eqref{w2} can be pure imaginary if the accessory parameter is considered to be pure imaginary. In spite of the QNFs \eqref{w3} and \eqref{w4}, whose real parts are indifferent to the size of the black hole,  the real part of $\omega$ in \eqref{w1} and \eqref{w2} increases as the black hole size $r_h$ increases.
Another difference is that, unlike the \eqref{w3} and \eqref{w4},  the real part of $\omega$ in \eqref{w1} and \eqref{w2} is inversely proportional to $k_\varphi$, which is consistent with the behavior of light orbital angular momentum for which the angular momentum is inversely proportional to the frequency   
\cite{yao2011orbital}.

\subsection{Quasinormal modes of Lifshitz black hole solution in  three-dimensional conformal gravity}\label{conqnm}

The Lifshitz black hole solution in three-dimensional conformal gravity has been described in section \ref{conformalgr}. 
Similar to the BTZ case, its event horizon is mapped into $X=Y=0$, and hence to construct the Dirac spinor $\Psi$ near the horizon, the solution for  $\Psi^R$ obtained in the previous subsection can be used.
In such a way that solution of the equation \eqref{kk},  gives $\chi_{10}(Y)$  by \eqref{xi1} that leads to
the $\Psi^R_{10}$  given by \eqref{Psi1}.
The $\Psi^R_{20}$ is also given by \eqref{psi22}, setting   $x=X$, $C_{13}=C_1$, and $C_{24}=C_2$. The behaviors of $\psi^R_1$ and $\psi^R_2$ on the horizon are also given by \eqref{horizon}.

However, as we explained before, despite the stringy black hole \eqref{mf}, whose range of the $\rho$ coordinate is incomplete, for the Lifshitz black hole solution \eqref{mconfo} the $\rho$  coordinate ranges in $(-\infty,+\infty)$, where the boundary at $r\rightarrow \infty$ corresponds to $\rho=\infty$, or equivalently $X=Y=\infty$, which is one of the regular singular points of the equation \eqref{kk}. 
There are certain transformations by which the solutions of the Heun equation at one singular point can be related to those at the other singular points \cite{Noda_2022,Suzuki_1998}. Here, we directly use the solution of the Heun differential equation at its regular singular point $Y=\infty$, where the solution of \eqref{kk} in terms of $X=-i\sqrt{2k}Y$ is given by \cite{abramowitz1968handbook}
\begin{equation}\label{xi1inf}
	\begin{split}
		\chi_{11}(X) =&{C_5}\,{X}^{-\alpha}{Hl}
		\left( -1, p_3,\alpha,\alpha_3,\mu_3,\delta,
		-i\sqrt{2k}X^{-1}\right)+{C_6}\,{X}^{-\beta
		}{Hl} \left( -1,p_4,\beta,
		\beta_3,\mu_4,\delta,-i\sqrt{2k}X^{-1} \right) ,
	\end{split}
\end{equation}
where $\epsilon=\beta+\alpha-\gamma-\delta+1$, $p_3=\left( \delta-\epsilon \right) \alpha+p$, $p_4= \left( \delta-\epsilon \right) \beta+p$, $\alpha_3=\alpha+1-
\gamma$, $\beta_3=
\beta+1-\gamma$, $\mu_3=\alpha-\beta+1$, and $\mu_4=\beta-\alpha+1$. Then, using \eqref{psi2}, the $\psi^R_1$ at infinity is given by
\begin{equation}\label{Psi1c}
	\psi^R_{11}(X)=\left(-i\sqrt{2k}+X\right)^{\xi } \left(i\sqrt{2k}+X\right)^{\nu } X^\sigma \chi_{11}(X).
\end{equation}
Also, substituting the above expression in \eqref{si2} gives the $\psi^R_2$ at infinity as follows 
\begin{equation}\label{xi2inf}
	\begin{split}
		\psi^R_{21}&(X)=\frac{\sqrt{k}   }{q\omega}\left(-i\sqrt{2k}+X\right)^{\xi } \left(i\sqrt{2k}+X\right)^{\nu } X^{-\xi+\nu}  \bigg( C_5X^{-\sigma}\big({\frac {iq\sqrt {2k}}{X}}Hlp\left( -1, p_3,\alpha,\alpha_3,\mu_3,\delta,-i\sqrt{2k}X^{-1}\right)\\
		&-Hl\left( -1, p_3,\alpha,\alpha_3,\mu_3,\delta,-i\sqrt{2k}X^{-1}\right)({X}^
		{2}+2\,k)^{-1}\big( iq
		\left( v-\xi \right) X\sqrt {2k}+q\sigma\,{X}^{2}+2\,k_{{
				\varphi}}X(\sqrt{k}   )^{-1}\\
		&+2\,qk \left( \sigma+\xi+v \right)  \big)\big)
		+C_6X^{\sigma}\big({\frac {iq\sqrt {2k}}{X}}Hlp\left( -1,p_4,\beta,
		\beta_3,\mu_4,\delta,-i\sqrt{2k}X^{-1} \right)\\
		&-Hl\left( -1,p_4,\beta,
		\beta_3,\mu_4,\delta,-i\sqrt{2k}X^{-1} \right)({X}^
		{2}+2\,k)^{-1}\big( iq \left( \nu-\xi \right) X\sqrt {2k}-q\sigma\,{X}^{2}+2\,k_{{
				\varphi}}X(\sqrt{k}   )^{-1}\\
		&-2\,qk \left( \sigma-v-\xi \right) 
		\big)
		\big)
		\bigg) .
	\end{split}
\end{equation}

Accordingly, at the spatial infinity, i.e. $\rho\rightarrow +\infty$, i.e $X\rightarrow +\infty$, we have
\begin{equation}\label{asym2}
	\begin{split}
		&\psi_{1\infty}^R= C_{5} {\rm e}^{{\frac{i\eta}{\sqrt{k}   }}\omega\rho}+C_{6} \,{\rm e}^{-{\frac{i\eta}{\sqrt{k}   }}\omega\rho},\\
		& \psi_{2\infty}^R =i\eta\left( -C_{5} {\rm e}^{{\frac{i\eta}{\sqrt{k}   }}\omega\rho}+C_{6} \,{\rm e}^{-{\frac{i\eta}{\sqrt{k}   }}\omega\rho}\right),
	\end{split}
\end{equation}
which show that with $\eta=1$ ($\eta=-1$) the second (first) terms denote {ingoing} waves.

In this black hole background, noting \eqref{Psi1}, \eqref{psi22}, \eqref{Psi1c}, and \eqref{xi2inf},  based on conformal relation \eqref{conf} in which the $h(\rho)$ function for this type of black hole is determined in \eqref{mfrr} and the $\Psi^R$ is defined as \eqref{eq:Solution}, the general form of solution for Dirac equation   can be constructed in the following form in terms of $X$
\begin{equation} \label{d2}
	\Psi(t,X,\varphi)=\frac{\sqrt{2}}{\sqrt{qX}}{\rm e}^{-i\omega t}{\rm e}^{ik_{\varphi}\varphi}\left(
	\begin{array}{cc}
		\psi_1^R(X) \\
		\psi_2^R(X)
	\end{array}
	\right),
\end{equation}
in which $q$ is fixed by \eqref{e} as $q=\frac{r_+}{\sqrt{2k}}$. It is worth reminding that the parameters of Heun functions are given by \eqref{alfabeta}-\eqref{constants}, and the complex energy levels are related to the  Heun functions parameters by \eqref{eq:31}. 

At the horizon, the wave function $\Psi^R$ behaves as \eqref{horizon}, where the energy flux at this boundary is again given by \eqref{fh}.  
Also,  the flux of energy at the boundary at spatial infinity, using \eqref{F} and \eqref{asym2}, reads
\begin{equation}\label{Fas2}
	{\cal F}_{asymp}\propto\eta
	\left( C_5C_5^*-C_6C_6^* \right) ,
\end{equation}
which admits that with $\eta=1$ ($\eta=-1$) the accompanied terms with $C_6$ ($C_5$) correspond to outgoing energy flux.  

At the horizon, similar to that of the BTZ black holes solution  discussed previously, imposing the QNMs boundary condition,  which requests  ingoing waves at this limit,
the $\eta=1$ case requires $C_{1}=0$ while the $\eta=-1$ case requires $C_{2}=0$ in \eqref{xi1}.
Also, at the spatial infinity, noting \eqref{asym2} and \eqref{Fas2}, in the $\eta=1$ case the first term in \eqref{xi1inf} should be excluded by setting $C_{5}=0$, while the $\eta=-1$ case requires $C_{6}=0$.  Then, having eliminated the ingoing parts of the wave function $\psi^R$ and the energy flux associated with $\Psi$, at the infinity the wave function $\Psi$ components determined by \eqref{d2} are proportional to $X^{-\frac{1}{2}- \frac{\omega_{Im}}{\sqrt{k}   }}$, 
using the relations \eqref{e}.
Accordingly,  unless $2\omega_{Im}>-\sqrt{k}$, or equivalently $2\omega_{Im}>-l^{-1}$ noting \eqref{e}, for which the wave function $\Psi$ vanishes at this boundary, only the outgoing modes are present at infinity.\footnote{At the event horizon, the wave function $\Psi$ is proportional to $X^{-\frac{1}{2}+ \frac{\omega_{Im}}{\sqrt{k}   }}$, that never vanishes with negative $\omega_{Im}$.}

Holding these conditions and then using \eqref{horizon} and \eqref{Fas2} in the last terms of the equation \eqref{stab2}, it can be concluded that the negative sign of the imaginary part of the $\omega$ is consistent with the stability condition of the Dirac equation solutions,  assuring that the integral in \eqref{stab2} is positive.

Now, for the wave function which admits the QNMs boundary conditions,  using the  parameters obtained in the previous section  \eqref{alfabeta}-\eqref{eq:31}, the QNFs can be rewritten in terms of parameters of black hole solution of conformal gravity, using \eqref{e}, as follows 
\begin{equation}\label{wc3}
	\omega 
	=\frac{1}{2\eta }\left(-\frac{1}{r_h}k_{\varphi}+\frac{i}{l}(1-p )\right)
	\quad{\rm with}\quad \kappa_{1}=-\kappa_{2}=-1,
\end{equation}
\begin{equation}\label{wc4}
	\omega 
	=\frac{1}{2\eta }\left(\frac{1}{r_h}k_{\varphi}+\frac{i}{l}(p +1)\right)
	\quad{\rm with}\quad \kappa_{1}=-\kappa_{2}=1.
\end{equation}
\begin{equation}\label{wc2}
	\omega 
	=\frac{1}{\eta l}\left(i-\frac {r_h}{2 l k_{{
				\varphi}}}p \right)
	\quad{\rm with}\quad \kappa_{1}=\kappa_{2}=-1,
\end{equation}
\begin{equation}\label{wc1}
	\omega 
	=\frac {r_h}{2 \eta l^2 k_{{
				\varphi}}}p 
	\quad
	{\rm with}\quad \kappa_{1}=\kappa_{2}=1,
\end{equation}
The QNFs are labeled by the Heun function's accessory parameter $p\in \mathbb{C}$. For larger black holes, the real part of $\omega$ in \eqref{wf1} and \eqref{wf2} grows, while it decreases in  \eqref{wf3} and \eqref{wf4} cases. Obviously, the QNFs \eqref{wc3}-\eqref{wc1} have the proper dimension of inverse length.
 Also, the stability of the QNMs  requires $p_{Re}>1$ ($p_{Re}<1$) for $\eta=1$ ($\eta=-1$) in \eqref{wc3}, $p_{Re}<-1$ ($p_{Re}>-1$) for $\eta=1$ ($\eta=-1$) in \eqref{wc4}, $\frac{\mathit{p_{Im}}}{k_{\varphi}}>\frac{2 l}{\mathit{rh}}$ ($\frac{\mathit{p_{Im}}}{k_{\varphi}}<\frac{2 l}{\mathit{rh}}$) for $\eta=1$ ($\eta=-1$) in \eqref{wc2}, and  $\frac{{p_{Im}}}{k_{\varphi}}<0$ ($\frac{{p_{Im}}}{k_{\varphi}}>0$) for $\eta=1$ ($\eta=-1$) in \eqref{wc1}.

In \cite{Gonzlez}, the propagation of massless fermionic fields in this black hole background has been considered by computing the  QNMs of the wave functions obtained in the form of hypergeometric functions. Their calculations lead to the QNFs of the form $\omega=\pm \frac{k_\varphi}{2r_h}-\frac{i}{2} (n+2)$. The obtained expressions in the cases of \eqref{wc3} and \eqref{wc4} can recast this form if one sets $\pm p =l\eta(n+2)+1$, where the positive and negative signs are for the combination \eqref{wc4} and \eqref{wc3}, respectively. Nevertheless, the QNFs \eqref{wc3}-\eqref{wc4} demonstrate more properly the inverse of the length dimension of $\omega$ with the dimensionless overtone number $p$.
In addition, the QNFs of the classes \eqref{wf1} and \eqref{wf2}  are not present in the QNFs group obtained in the procedure based on hypergeometric functions, provided in \cite{Gonzlez}.

\subsection{Quasinormal modes of Lifshitz exact solutions of Hu-Sawicki $F(R)$ model with 	Hyperscaling violation}

The  Lifshitz exact solutions of Hu-Sawicki $F(R)$ model with 	Hyperscaling violation
has been described in section \ref{Sawicki}.
Applying the same procedure   based on the Weyl symmetry of the Dirac action that leads to \eqref{conf}, where the $h(\rho)$ function for this type of black hole can be read from \eqref{mfrr}, the solution for Dirac equation in this background is constructed in the form of
\begin{equation} \label{d3}
	\Psi(t,X,\varphi)=\frac{{\sqrt{q(X^2+2k)}}}{\sqrt{X}}{\rm e}^{-i\omega t}{\rm e}^{ik_{\varphi}\varphi}\left(
	\begin{array}{cc}
		\psi_1^R(X) \\
		\psi_2^R(X)
	\end{array}
	\right),
\end{equation}
in which 
the $q$ parameter is fixed here by \eqref{e1} as 
$
q={\frac {\sqrt {2}{l}^{z}{{r_h}}^{1-\frac{z}{2}}}{2\sqrt {ak}}
}$. 
As said before, for this class of black hole solutions the range of the $\rho$ coordinate is complete. The $\psi_1^R(X)$ and $\psi_2^R(X)$ obtained for the black hole solution in three-dimensional conformal gravity, discussed in section \ref{conqnm}, can be used for this background.
In such a way that at the event horizon, similar to those of the other two black hole backgrounds, 
the solution of \eqref{kk}  gives $\chi_{10}(Y)$  by \eqref{xi1} that leads to the $\Psi^R_{10}$  given by \eqref{Psi1}, while the $\Psi^R_{20}$ is given by \eqref{psi22}, setting   $x=X$, $C_{13}=C_1$, and $C_{24}=C_2$. The behavior of $\psi^R_1$ and $\psi^R_2$ on the horizon is also given by \eqref{horizon}. 
At spatial infinity where $\rho\rightarrow \infty$, the solutions for the $\psi_1^R$ and $\psi^R_2$ are given by \eqref{Psi1c} and \eqref{xi2inf}, whose asymptotic behaviors are given by \eqref{asym2}. The energy fluxes at the horizon and spatial infinity are also given by \eqref{fh} and \eqref{Fas2}.

Similar to the fermionic perturbation in Lifshitz black hole solution in conformal gravity, the QNMs boundary conditions in this black hole background require $C_{1}=C_5=0$  for  $\eta=1$ case, and $C_{2}=C_6=0$ for $\eta=-1$ case. 
Holding these conditions, the wave function $\Psi$ components determined  by \eqref{d3} at they horizon are proportional to ${\rm e}^{\left(-\frac{1}{2}+ \frac{\omega_{Im}}{\sqrt{k}   }\right)\rho}$, while at the spatial infinity the are  proportional to ${\rm e}^{\left(\frac{1}{2}- \frac{\omega_{Im}}{\sqrt{k}   }\right)\rho}$.  It shows that, with negative $\omega_{Im}$, required for stability of the solutions, the wave function $\Psi$  does not vanish at any of the boundaries.

In terms of the model parameters and radius of horizon determined by \eqref{e1}, using  the  parameters   \eqref{alfabeta}-\eqref{eq:31}, the QNFs \eqref{eq:31} recasts the following forms for different combinations of the $\kappa_{1}$ and $\kappa_{2}$
\begin{equation}\label{wf3}
	\begin{split}
		\omega
		={\frac { {{ r_h }}^{z-2}{l}^{1-z}\sqrt { R_0 }}{8
				\left( z-2 \right) \eta }}&\left( -2\sqrt {2}k_{{\varphi}}+i\sqrt { R_0 } \left( 
		1-p \right)  \right)
		, \quad{\rm with}\quad \kappa_{1}=-\kappa_{2}=-1,
	\end{split}
\end{equation}
\begin{equation}\label{wf4}
	\begin{split}
		\omega
		={\frac {\sqrt { R_0 }{{ r_h }}^{z-2}  {l}^{1-z}}{8
				\left( z-2 \right) \eta }}&\left( 2\,\sqrt {2}k_{{
				\varphi}}+i\sqrt { R_0 } \left( 1+p \right)  \right)
		, \quad{\rm with}\quad \kappa_{1}=-\kappa_{2}=1,
	\end{split}
\end{equation}
\begin{equation}\label{wf2}
	\omega
	={\frac {{ R_0 }^{\frac{3}{2}}\sqrt {2}{l}^{1-z}{{ r_h }}^{z-2}}{32
			\left( z-2 \right) \eta } \left( 4\,i\sqrt {2}-{\frac {\sqrt {{ R_0
			}}}{k_{{\varphi}}}}p \right) }, \quad{\rm with}\quad \kappa_{1}=\kappa_{2}=-1,
\end{equation}
\begin{equation}\label{wf1}
	\omega
	={\frac {{{R_0}}^{\frac{3}{2}}{{ r_h }}^{z-2}{l}^{1-z}\sqrt {2}}{32
			\left( z-2 \right) k_{{\varphi}}\eta}}p \quad{\rm with}\quad \kappa_{1}=\kappa_{2}=1
	.
\end{equation}
Similar to those of the other considered black holes, the imaginary part of $\omega$ in \eqref{wf1} and \eqref{wf2} is labeled by the imaginary part of the accessory parameter $p$, while for \eqref{wf3} and \eqref{wf4} it is labeled by the real part of $p$. The QNFs in the classes of \eqref{wf1} and \eqref{wf2} can be pure imaginary if the $p$ constant is considered to be imaginary in \eqref{wf1}  and zero in  \eqref{wf2}.
The stability of the QNMs with $z>2$ requires $p_{Re}>1$ ($p_{Re}<1$) for $\eta=1$ ($\eta=-1$) in \eqref{wf3}, $p_{Re}<-1$ ($p_{Re}>-1$) for $\eta=1$ ($\eta=-1$) in \eqref{wf4}, $\frac{{p_{Im}}}{k_{\varphi}}>\frac{4 \sqrt{2}}{\sqrt{{R_0}}}$ ($\frac{{p_{Im}}}{k_{\varphi}}<\frac{4 \sqrt{2}}{\sqrt{{R_0}}}$) for $\eta=1$ ($\eta=-1$) in \eqref{wf2}, and  $\frac{{p_{Im}}}{k_{\varphi}}<0$ ($\frac{{p_{Im}}}{k_{\varphi}}>0$) for $\eta=1$ ($\eta=1$) in \eqref{wf1}.
\section{Conclusion}\label{conclussion}

Implementing a family of special functions: Heun functions, we studied fermionic field perturbations of three black hole solutions in $2+1$-dimensional gravity, including the BTZ black hole solution, the Lifshitz black hole solution in three-dimensional conformal gravity obtained in \cite{conformalbh}, and the Lifshitz exact solutions in Hu-Sawicki $F(R)$ model with 	Hyperscaling violation obtained in \cite{hendi}. The interesting shared feature by these black holes is the Weyl relation of their line element to that of a homogeneous spacetime, whose spatial part possesses the symmetry of two-dimensional Lie Algebra.
Hence, the local Weyl symmetry of the massless Dirac action under the conformal transformation of metric connects the Dirac spinors in these three backgrounds to those of the homogeneous spacetime. It enabled us to calculate analytically the QNFs of massless fermionic propagating outside the event horizon of these black holes by solving the set of differential equations on the homogeneous spacetime. The wave functions on the black hole backgrounds are then obtained by  Weyl transformations. 
The solutions to the Dirac equation on the homogeneous spacetime are obtained in terms of the Heun functions, which are not reducible to the hypergeometric functions.

The solutions of the wave equations at the three black hole backgrounds are obtained to be characterized by complex frequencies. 
On the event horizon, the QNMs boundary condition, which requires pure ingoing waves, is imposed in a similar way on the obtained solutions for all considered backgrounds.  At spatial infinity, on the other hand,  the associated conformal factor for the BTZ black hole required the wave function 
to be dealt with in a different way from those of the other two backgrounds. The reason is that for the BTZ black hole solution the region $r_h \leq r < \infty$  amounts to $-\infty < \rho \leq\rho_0$, which is not a complete range, similar to those of asymptotically anti-de Sitter spacetimes \cite{Konoplya}. Hence,  the boundary condition $\Psi\rightarrow 0$ at $r\rightarrow \infty$, prescribed for the cases that the coordinate has upper bound \cite{Kodama_2004} is adopted for BTZ background, while for the other two backgrounds, the existence of pure outgoing waves at infinity is demanded.

It has been argued that to have stable propagation of the fermionic field in these three backgrounds,   the imaginary part of the QNFs should be always negative.  The QNFs are expressed in terms of the black hole parameters and labeled by the 
accessory parameter $p\in \mathbb{C}$, inherited from the Heun function. We classified the QNFs of each background into four groups. It has been shown that with a specific set of $p$ parameter in the  BTZ black hole, two groups of the obtained QNFs via the Heun function can recast a similar form to those calculated via hypergeometric functions in \cite{BTZQN,btzqnmhyper} and Heun function in \cite{Kwon_2011}.
Nevertheless,   in addition to these two groups, our calculations based on the Heun function bring up two other groups of QNFs, which have not been present in the spectrum provided by the other approaches based on hypergeometric and Heun functions.
Similarly, two groups of QNFs obtained for the  Lifshitz black hole solution in conformal gravity can recast the form of QNMs calculated for this background using the hypergeometric function in  \cite{Gonzlez}, 
when the accessory parameter of the Heun function is expressed in terms of the overtone number $n$.  In addition to these two groups of QNFs, which can be reduced to the known sequences of QNFs, the calculations based on the Heun function bring up two other groups of QNFs, that have not been present in the spectrum provided by the hypergeometric function. 
We have also obtained the QNMs and the associated QNFs for the Lifshitz exact solutions of the Hu-Sawicki $F(R)$ model. Imposing the boundary conditions, the stability condition is also considered. The QNFs of this background have been expressed in terms of the black hole parameters and the accessory parameter $p$.

One of the differences between the aforementioned two cases of QNFs appeared particularly in the Heun function formalism with those two cases which are also present in the hypergeometric function approach is that the real part of the new QNFs are inversely proportional to the orbital angular momentum quantum number $k_\varphi$, in spite of the other two QNFs which are directly proportional to $k_\varphi$. The inverse relation of frequency and the orbital angular momentum is a particular characteristic of light beams
with an azimuthal phase dependence of exp$(ik_\varphi \varphi) $ \cite{yao2011orbital}. In this context, it is worth investigating the implications of the obtained wave functions and their associated QNMs in physical systems.

 

	
\bibliographystyle{h-physrev}
\bibliography{blackhole22}
\end{document}